%% file: iclr2025_conference.tex
\documentclass{article} 
\usepackage{graphicx}
\usepackage{AI4NA_workshop_2025,times}
\usepackage{algorithm}
\usepackage{algorithmic}
\usepackage{booktabs} 
\usepackage{multirow}

\iclrfinalcopy 

\input{math_commands.tex}

\usepackage{hyperref}
\usepackage{url}

\title{ RiboGen: RNA Sequence and Structure Co-Generation  with Equivariant MultiFlow}

\author{%
  Dana Rubin \\
  MIT CSAIL \\
  MIT Media Lab, Molecular Machines \\
  \texttt{danaru@mit.edu} \\
  \And
  Allan dos Santos Costa \\
  Center for Bits and Atoms\\
  MIT Media Lab, Molecular Machines\\
  \texttt{allanc@mit.edu} \\
  \And
  Manvitha Ponnapati \\
  Center for Bits and Atoms\\
  MIT Media Lab, Molecular Machines\\
  \texttt{} \\  
  \And
  Joseph Jacobson \\
  Center for Bits and Atoms\\
  MIT Media Lab, Molecular Machines\\
  \texttt{} \\  
}

\iclrfinalcopy

\begin{document}

\maketitle

\begin{abstract}
Ribonucleic acid (RNA) plays fundamental roles in biological systems, from carrying genetic information to performing enzymatic function. Understanding and designing RNA can enable novel therapeutic application and biotechnological innovation. To enhance RNA design, in this paper we introduce RiboGen, the first deep learning model to simultaneously generate RNA sequence and all-atom 3D structure. RiboGen leverages the standard Flow Matching with Discrete Flow Matching in a multimodal data representation. RiboGen is based on Euclidean Equivariant neural networks for efficiently processing and learning three-dimensional geometry. Our experiments show that RiboGen can efficiently generate chemically plausible and self-consistent RNA samples, suggesting that co-generation of sequence and structure is a competitive approach for modeling RNA.
\end{abstract}

\section{Introduction}
Ribonucleic acid (RNA) is a fundamental biomolecule that stands at the intersection of modern biology and the origins of life. RNA has proven to be a versatile molecule, playing key roles in messaging \citep{crick1970centraldogma}, catalytic functions \citep{guerrier1983rna}, regulation, and diverse biological processes through its complex 3D structures \citep{fire1998rna}. While traditional computational methods faced limitations in decoding RNA structure and facilitating RNA design, deep learning emerged as powerful approach to accurately predict RNA structures, enhance RNA engineering, and unlock new insights into its functional roles. Existing deep learning models often predict RNA structure from sequence or design sequences for target structures independently. However, the ability to simultaneously generate both sequence and structure remains a largely unexplored space in deep learning-based modeling of RNA. This co-generation capability can enable exploration of the sequence-structure landscape in novel ways. To address this gap, in this paper we introduce RiboGen for joint generation of all-atom structure and sequence of RNA. RiboGen is based on a Multiflow \citep{campbell2024generative} model that draws on Flow Matching \citep{lipman2022flow, liu2022flow} and Discrete Flow \citep{gat2024discreteflowmatching, campbell2024generative} for its generation. We train a large model and evaluate its chemical validity and self-consistency. Our results showcase the capabilities of RiboGen for generation and highlight that co-generation models are a promising avenue for modeling of RNA.

\subsection{Related Work}
Previous work in deep learning for RNA design has made significant progress in computational approaches to predict RNA structures \citep{shen2024accurate, abramson2024}, RNA-RNA interactions, and the design of novel RNA sequences. Recent progress in generative modeling for RNA focuses on sequence and structure generation through Denoising Diffusion Probabilistic Models (DDPM) \citep{ho2020denoising} or Flow Matching \citep{lipman2022flow}. MMDiff \citep{morehead2023jointsequencestructuregenerationnucleic} uses discrete DDPM to co-generate sequence and structure of RNA, DNA and proteins. Our approach instead employs Flow Matching and its discrete variant \citep{campbell2024generative}. RNA-FrameFlow 
\citep{anand2024rna} represents RNA through rigid body frames and uses flow matching to generate 3D backbones, employing inverse folding model gRNAde \citep{joshi2023grnade} to obtain sequences. In contrast, while similarly using Flow Matching for 3D generation, our approach additionally models the discrete sequence components of RNA generation. RNAFlow \cite{nori2024rnaflow} uses a GNN conditioned on protein structure and sequence to generate RNA sequences, which are then processed by RoseTTAFold2NA \cite{Baek2024} to predict backbone structure; this method additionally conditions on protein structure as input. Our approach instead focuses on isolated RNA, learning unconditional direct sequence-structure generation. Recent application of the Multiflow \citep{campbell2024generative} framework to protein sequence-structure design has demonstrated the power of joint generation. Our approach builds upon on these insights from protein design, and adapts them to the RNA design domain by enabling the joint generation of RNA sequences and all-atom structures.


\begin{figure}[t]
    \centering
    \includegraphics[width=1.0\linewidth]{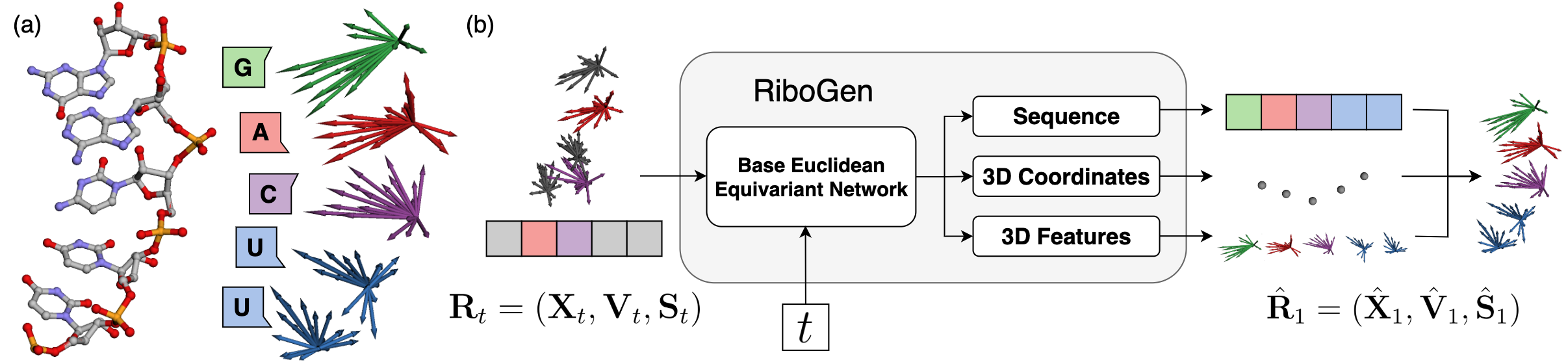}
    \caption{\textbf{RNA Sequence and Structure Co-Generation}: (a) Traditional molecular structure showing the nucleotides with atoms and bonds. Right side demonstrates how each nucleotide (G, A, C, U) is represented as both a discrete sequence element (colored boxes) and associated 3D point cloud representation (colored directional features) centered around the $C3'$ atom. (b) The RiboGen model architechture: the model takes noised input of sequence and geometric features $\mathbf R_t$, and a time parameter $t$, process them through the base network and simultaneously predicts three components: the RNA sequence, central coordinates, and 3D features. These components are combined to produce the final RNA structure prediction $\hat{\mathbf R}_1$. }
    \label{fig:model}
\end{figure}

\section{Methods}

\subsection{RNA Representation}
We represent an RNA molecule as a sequence and a 3D gas of geometric features $\mathbf R = (\mathbf S, \mathbf X, \mathbf V)$ (Figure \ref{fig:model}.a), where: 
\begin{itemize}
    \item \( \mathbf S \in \mathcal{S}^N \) is sequence of length \( N \), formed out of standard nucleotides \( \mathcal{S} = \{A, C, G, U\} \).
    \item \( \mathbf X \in \mathbb{R}^{N \times 3} \) contains the 3D coordinates of the C3' atom for each nucleotide, chosen as the \textbf{reference center}.
    \item \( \mathbf V \in \mathbb{R}^{N \times 24 \times 3} \) are geometric features encoding the relative position of up to 24 heavy atoms per nucleotide to its center at C3', in canonical ordering. This representation encompasses both the sugar-phosphate backbone atoms and the base atoms, allowing for complete reconstruction of the RNA structure. Nucleotides that have fewer than 24 heavy atoms have their corresponding channels of $\mathbf V$ padded with zeros.  
\end{itemize}

After generating the three components, the predicted vectors \( \mathbf V \) are added to the predicted centers \(  \mathbf X \) and the sequence $\mathbf S$ is utilized for labeling nucleotides  and atomic types, for full reconstruction of 3D atomic coordinates. This representation encodes both the chemical identity and geometry of each nucleotide while preserving rotational and translational equivariance, which is essential for downstream learning via Euclidean Equivariant Neural Networks.

\subsection{Flow Matching}

To model the distribution of 3D RNA coordinates $\mathbf X$ and features $\mathbf V$ we use Flow Matching \citep{lipman2022flow, liu2022flow, albergo2023stochastic}. Flow Matching  parameterizes a conditional probability path $\rho_t(\mathbf X_t |\mathbf X_1)$ on time $t$ by learning a conditional velocity field $\hat v^\theta_t(\mathbf X_t) \approx v_t(\mathbf X_t|\mathbf X_1)$ that transforms samples from a prior distribution $\mathbf X_0 \sim \rho_0 =\mathcal N$ to a target data distribution $\mathbf X_1 \sim \rho_1 = \rho_D$. To learn this transport, we use the standard form of Flow Matching to obtain a noised version of $\mathbf X_1$ via the linear interpolant and its associated velocity:
\begin{equation}
\mathbf X_t = (1-t) \mathbf X_0 + t \mathbf X_1
\end{equation}
\begin{equation}
v_t(\mathbf X_t|\mathbf X_1) = \mathbf X_1 - \mathbf X_0
\end{equation}
We build our model to reconstruct the target $\hat{\mathbf X}^\theta_{1|\mathbf X_t} \approx \mathbf X_1$ from its noised counterpart $\mathbf X_t$. We follow the reparameterization of \citep{jing2024alphafold, pooladian2023multisampleflowmatchingstraightening} and obtain the learned conditional velocity through:
\begin{equation}
v_t^\theta(\mathbf X_t) =  \frac{1}{(1-t)} \Big (\hat{\mathbf X}^\theta_{1|\mathbf X_t}- \mathbf X_t \Big ) \approx v_t (\mathbf X_t | \mathbf X_1)
\label{eq:learned-velocity-continuous}
\end{equation}
We then sample our learned model via integration $\mathbf X_1 = \mathbf X_0 + \int_{0}^1  v_t^\theta(\mathbf X_t) dt$ where $\mathbf X_0 \sim \mathcal N$.

\subsection{Discrete Flow Matching}

While the standard form of Flow Matching is effective for continuous data, it is not appropriate for categorical domains. Hence, for modeling the RNA sequence we employ the extended framework of Discrete Flow Matching \citep{gat2024discreteflowmatching, campbell2024generative}. In this setting, the sequence data $\mathbf S\in\mathcal S^N$ is described over a vocabulary $\mathcal S$. We parameterize a discrete flow by describing the velocity field over a probability vector on this categorical space:
\begin{equation}
\mathbf S_t \sim \textrm{Cat}((1-t)\delta_{\mathbf S_0} + t \delta_{\mathbf S_1})   
\end{equation}
\begin{equation}
v_t(\mathbf S_t|\mathbf S_1) = \delta_{\mathbf S_1} -\delta_{\mathbf S_0}
\end{equation}
where $\delta_\mathbf S \in \mathbb R^{N\times|\mathcal X|}$ is the Dirac delta representation of $\mathbf S$ and Cat($\cdot$) denotes the categorical distribution. We learn a model to predict the probability vector $p^\theta_{1|\mathbf S_t} \approx \delta_{\mathbf S_1}$. In similar reparameterization to Equation \ref{eq:learned-velocity-continuous}, we obtain the approximate conditional velocity through:
\begin{equation}
v^\theta_t(\mathbf S_t) = \frac{1}{(1-t)} \Big  (\hat p^\theta_{1|\mathbf S_t} -\delta_{\mathbf S_t} \Big ) \approx v_t(\mathbf S_t|\mathbf S_1) 
\end{equation}

\begin{figure}[t]
    \centering
    \includegraphics[width=1.0\linewidth]{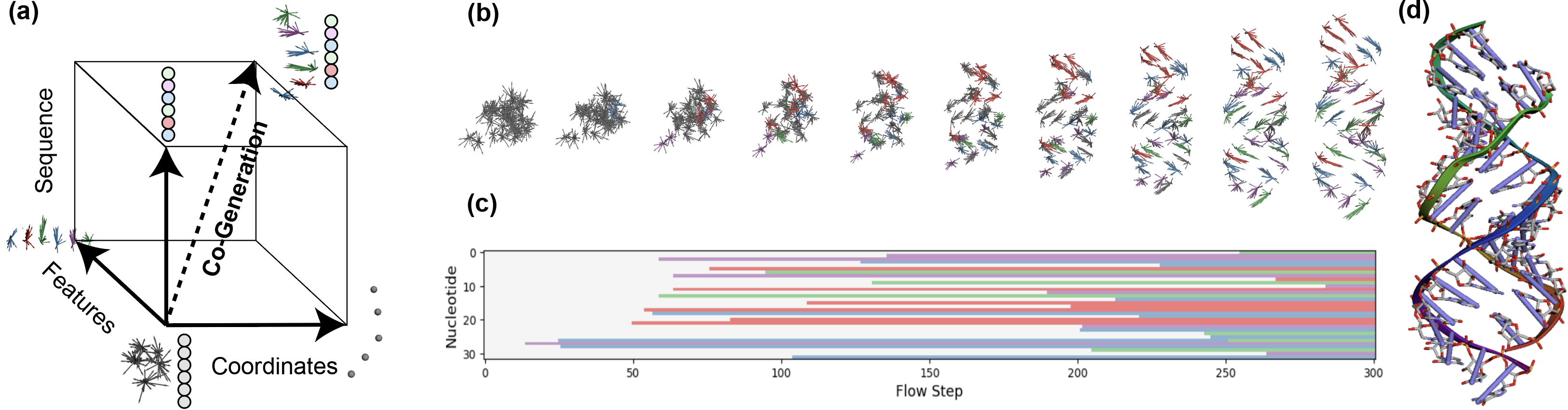}
    \caption{\textbf{Multiflow for RNA Sequence, Backbone and Atomistic Structure}: (a) Schematic representation of our Multiflow approach, demonstrating the three dimensions- sequence, coordinates, and features. (b) Visualization of the  RNA structure generation across multiple time steps. (c) Visualization of the Discrete flow matching used for sequence prediction in the model, where each color represents a different nucleotide. (d) Final product, a complete generated RNA molecule.}
    \label{fig:TC}
\end{figure}

\subsection{Multiflow}
To generate full RNA representations, we use Multiflow \citep{campbell2024generative} and train a neural network to learn the multimodal velocity field $d\mathbf R_t=(d\mathbf S_t, d\mathbf X_t, d\mathbf V_t)$ given jointly noised data $\mathbf R_t$ and time $t$ (Figure \ref{fig:model}.b). We employ the standard Flow Matching for $\mathbf X$ and $\mathbf V$ and its discrete counterpart for $\mathbf S$. This decomposition enables the model to separately capture the distributions of sequence, backbone, and atomic positions, allowing for unconditional generation or conditional generation based on specific structural or sequence constraints (Figure \ref{fig:TC}.a), such as structure prediction or inverse folding.

\subsection{Architecture and Training}
We use Euclidean-Equivariant Neural Networks \citep{geiger2022e3nn} for processing our RNA representation. To handle the different modalities of $\mathbf R$, our model consists of a base network that feeds into 3 headers for each data component: sequence, coordinates and 3D features. The sequence header predicts a probability vector $\hat p_{\mathbf R_1} \in \mathbb R^{N\times |\mathcal S|}$, while coordinate and feature headers predict equivariant variables $\hat {\mathbf {X}}_1$ and $\hat {\mathbf {V}}_1$. We train the model to reconstruct the original structure $(\mathbf X, \mathbf V)$ and sequence $\mathbf S$:
\begin{align}
\mathcal L = \mathcal L_{\textrm{struct}} + \mathcal L_{\textrm{seq}} = \textrm{Mean}(\|V(\mathbf X_1,\mathbf V_1) - V(\hat {\mathbf X}_{1},\hat{\mathbf V}_{1})\|^2) + \textrm{CrossEntropy}(\hat p_{\mathbf S_1}, \delta_{\mathbf S_1})
\end{align}
where $V(\mathbf X, \mathbf V) \in\mathbb R^{N_A \times N_A \times 3}$ is the 3D vector map between every atom in the system (size $N_A$). 

\section{Results}

We leverage the RNASolo dataset \citep{adamczyk2022rnasolo} which consists of extracted individual RNA structures from the Protein Databank (PDB) \citep{berman2000protein} to train our model. Following \citep{anand2024rna}, we filter the full dataset to resolution $< 4 \mathring{\text{A}}$ and sequence lengths between 40 and 150 to a total dataset size of 6090 data points. The dataset exhibits significant length imbalances which led to biased model performance initially, to address this, we implemented a length-balanced sampling that ensures uniform representation of RNA sequences for training as described in Appendix~\ref{subsec:data_balancing}.
We train our model with batch size of 64 for 120k steps on 4 GPUs. 
Our model's flow process is trained on 300 timesteps. 
To evaluate our model, we follow \citep{anand2024rna} and sample 50 RNA structures at each sequence length from 40-150 with step size of 10.

\begin{figure}[h!]
    \centering
    \makebox[\textwidth][c]{%
    \includegraphics[width=1\linewidth]{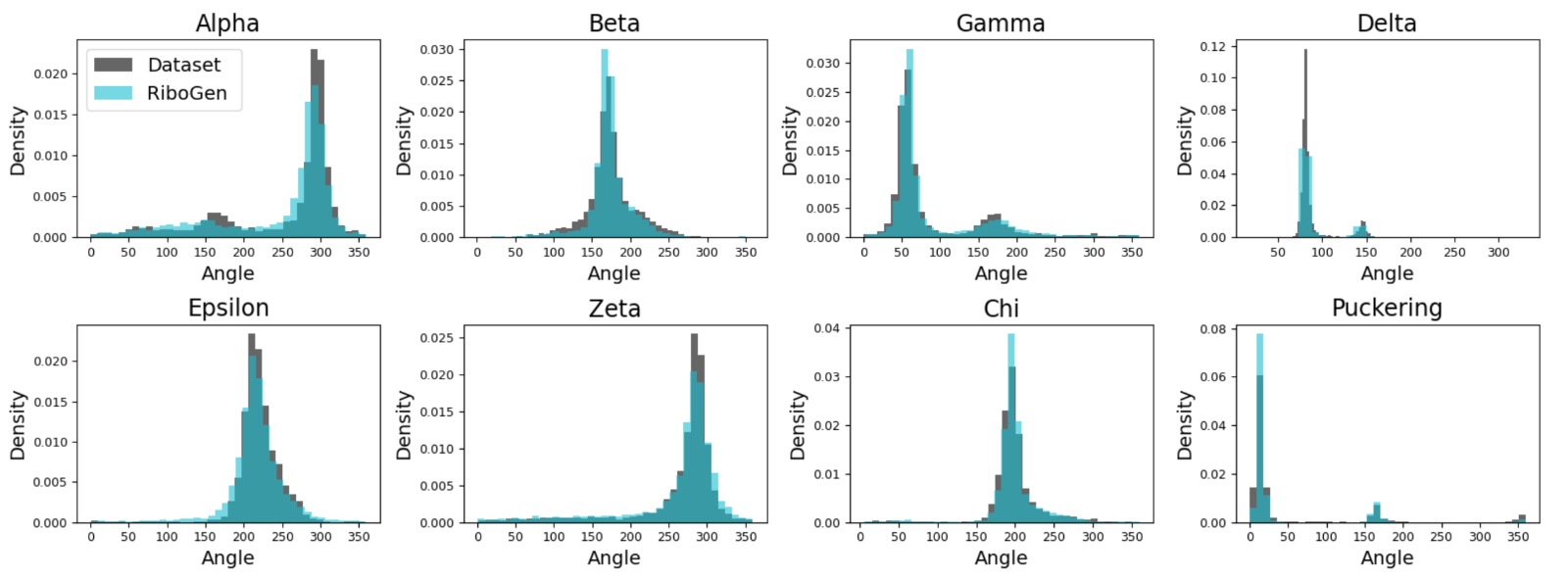}
    }
    \caption{\textbf{RiboGen Chemical Analysis}: Distribution comparison of key RNA geometric parameters between the training dataset and 50 random samples of RiboGen generations across all lengths. (5 from each length) The analyzed parameters include alpha, beta, gamma, chi dihedral angles, and ribose puckering phase, which are strong indicators of RNA backbone and chemical validity.}
    \label{fig:chem2}
\end{figure}

\subsection{Chemical Validity}
To assess the chemical validity of our generated structures, we analyzed key geometric parameters that define RNA backbone and base conformations. Using MDAnalysis \citep{Gowers2016} \citep{Michaud-Agrawal2011} we computed all dihedral angles and the pseudo-angle for the ribose pucker. Figure \ref{fig:chem2} shows the distributions of the dihedral angles across the training data (representing experimentally determined structures) and 50 randomly selected RiboGen structures, 5 from each sequence length. The results demonstrate that for most angles RiboGen generated structures capture the dihedral angles distributions and the general trends of the training set, though with some discrepancies. Our RiboGen structures show broader distribution across the Alpha angle indicating mild divergence from the experimental data in this specific torsion. 
Overall these results suggest that RiboGen has successfully learned the geometric constraints of RNA molecules.

\begin{figure}[t]
    \centering
    \includegraphics[width=0.8\linewidth]{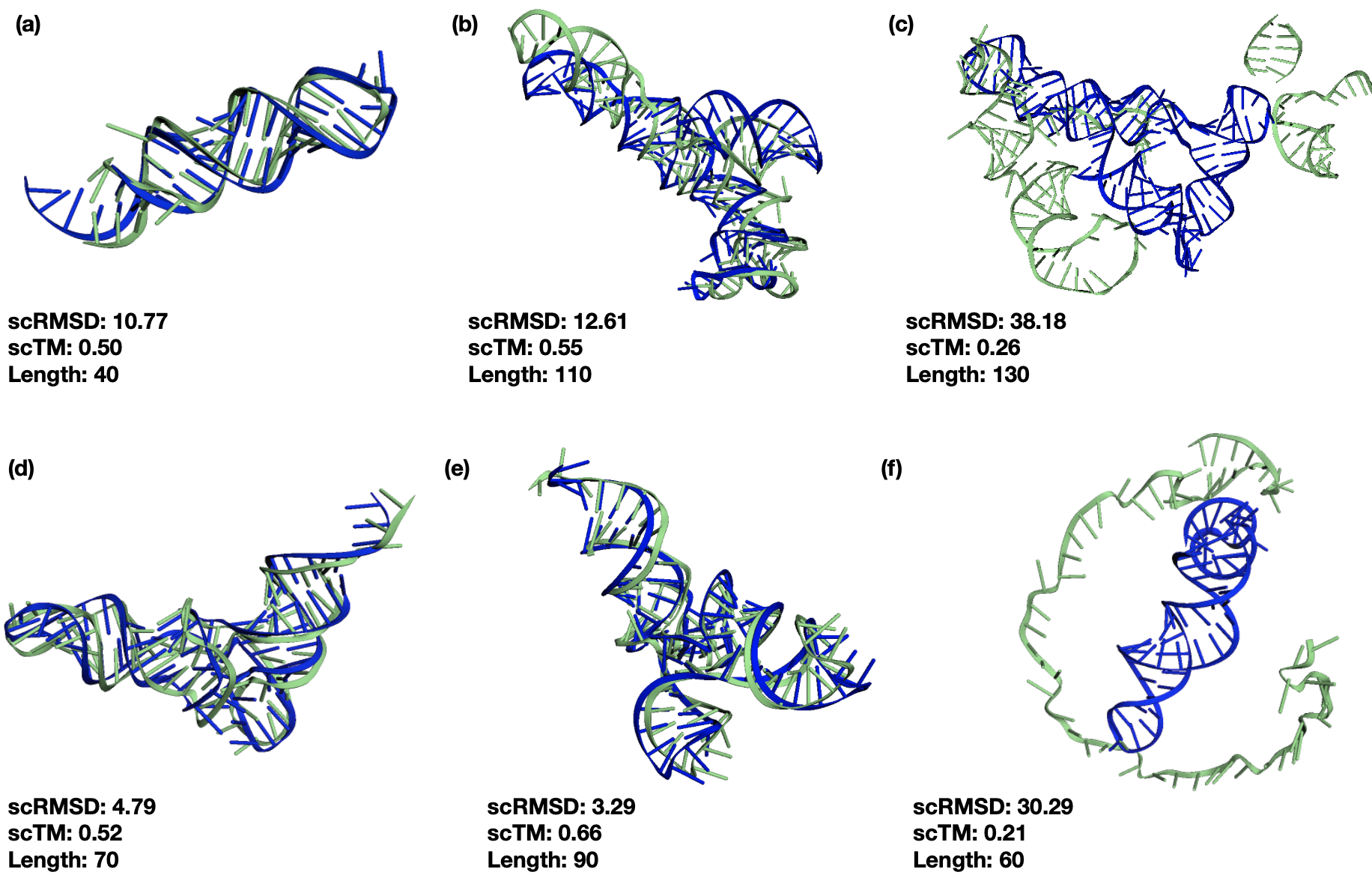}
    \caption{\textbf{Self-consistency Visualization of RiboGen's Joint Sequence-Structure Generation Aligned with Boltz Structure:}
     RiboGen-generated RNA structures (green) aligned with Boltz structure predictions (blue) derived from the corresponding co-generated sequences of RiboGen. Six examples across different sequence lengths demonstrate varying degrees of structural agreement.
     Notably, in some cases (c, f) RiboGen generates fragmented or unfolded structures, suggesting failure modes in the sampling process for long or structurally complex sequences.}
    \label{fig:align_samples}
\end{figure}

\begin{figure}[t]

    \centering
        \makebox[\textwidth][c]{%
        \includegraphics[width=1\linewidth]{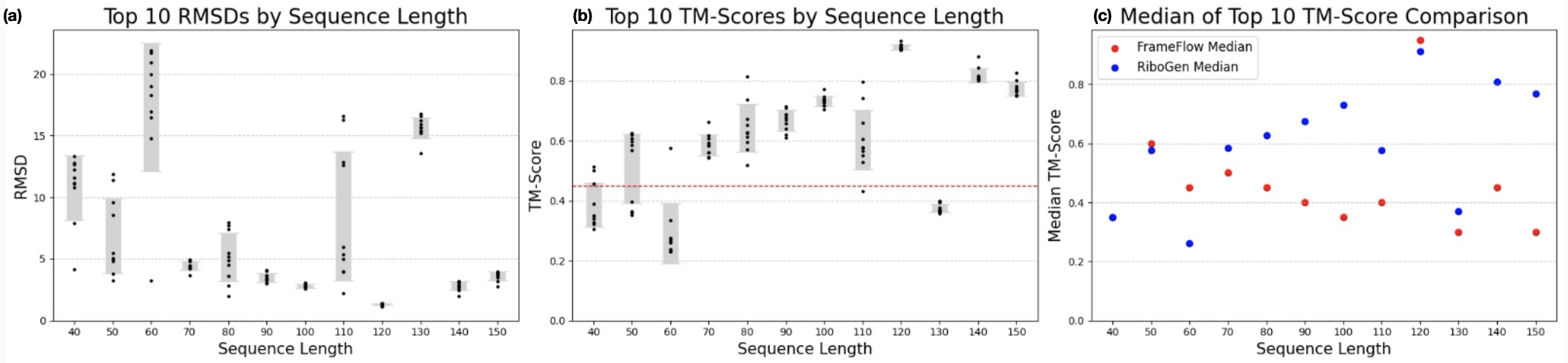}
    }
    \caption{\textbf{Self-Consistency Evaluation}: (a) RMSD and (b) TM-score between our generated structures and Boltz-1 predictions across different sequence lengths (40-150 nucleotides), showing the top 10 generated structures for each length. The TM-score ranges from 0 to 1, with higher values indicating better structural agreement, while lower RMSD values indicate better structural similarity. (c) Median of TM-scores of top 10 generated structures: The plot compares RiboGen's and FrameFlow's medians across various RNA sequence lengths and illustrates that RiboGen achieves higher TM-scores for RNA sequences between 70-150 nucleotides, excluding 120 which has similar median. FrameFlow demonstrates comparable performance for shorter sequences but shows decreased structural accuracy as sequence length increases. }
    \label{fig:TMscore}
\end{figure}

\subsection{Self-Consistency}\label{sec:sc}
To evaluate the quality and biological plausibility of our generated RNA, we employed a self consistency validation process. For each generated RNA molecule, we extracted its sequence and used Boltz-1 \citep{boltz1} to obtain a reference structure. We quantify structural similarity between our generated structure and Boltz's structure using two complementary metrics: Root Mean Square Deviation (RMSD) and Template Modeling score (TM-score). We calculated these metrics across all 50 samples for each sequence length. Following \citep{anand2024rna}, in Figure \ref{fig:TMscore}(a) and (b) we report results on 10 best-performing samples per length. Our predictions scTM scores in \ref{fig:TMscore}(b) demonstrate lower variance than in \cite{anand2024rna}, suggesting better consistency and generalization across different lengths.
In Figure \ref{fig:TMscore}(c) we observe that RiboGen achieves higher median TM-scores than RNA-FrameFlow for most sequence lengths above 70 nucleotides. This  may highlight the benefits of co-generation, especially for longer RNA.

\subsection{Structural Evaluation}\label{sec:structure}
To evaluate RiboGen's performance in generating valid RNA structure-sequence pairs, we utilized the metrics from the evaluation suite proposed by \citep{anand2024rna}. While RNA-FrameFlow focuses on structure generation alone, RiboGen jointly generates both the sequence and its corresponding structure.  Therefore, instead of using gRNAde \citep{joshi2025grnadegeometricdeeplearning} sequences and corresponding RhoFold \citep{shen2024accurate} structures to calculate TM-score, we folded the co-generated sequences using Boltz-1 and computed the TM-score after aligning them to the generated structures. While RNA-FrameFlow employs an additional 8-shot inverse folding process when calling gRNAde eight times for each backbone, RiboGen performs one-shot co-generation of both sequence and structure in a single sampling process. This highlights RiboGen’s potential for simpler and more efficient RNA design workflows, avoiding the need for expensive post-hoc sequence inference. Following \citep{anand2024rna}, samples with $\text{TM-score} \geq 0.45$ were considered valid. To assess diversity, we measured the number of unique qTM clusters among the valid samples and normalized this by the total number of valid samples. Although RiboGen is sampling jointly RNA sequences and structures, it achieved performance on par with RNA-FrameFlow in terms of backbone validity. Our results demonstrate competitive metrics and efficient sampling, validating RiboGen as a promising baseline for joint RNA structure-sequence generation.

\begin{table}[h!]
\centering

\begin{tabular}{lcccc}
\toprule
\textbf{Model} & \textbf{Sampling Steps $N_T$} & \textbf{\% Validity $\uparrow$}  & \textbf{Time (s) $\downarrow$} \\
\midrule
\multirow{3}{*}{RiboGen}
 & 100 & 27.17 &  \textbf{1.18} \\
 & 200 & 32.17 &  4.50 \\
 & 300 & \textbf{34.17} & 9.06 \\
\midrule
\multirow{3}{*}{RNA-FrameFlow * } 
& 10 & 16.7 &  --   \\
 & 50 & \textbf{41.0} &   4.74   \\
 & 100 & 20.0 &   --   \\
\midrule
MMDiff * &  100  &  0      & 27.30 \\

\bottomrule
\end{tabular}
\caption{\textbf{Performance comparison of unconditional RNA structure generation models}.
This table presents RiboGen's performance across varying flow sampling timesteps ($N_T$), alongside other models. Metrics include structural validity, and computational cost in seconds per generation. * Results for RNA-FrameFlow and MMDiff \citep{morehead2023mmdiff}
 methods are reported from \citep{anand2024rna}.}
\end{table}

\section{Conclusion}
In this paper, we introduced RiboGen, the first generative model to jointly produce RNA sequences and their corresponding all-atom 3D structures by learning a single multi-modal Flow field. Our approach leverages Flow Matching for continuous structural components and Discrete Flow Matching for sequence generation within the Multiflow framework. We demonstrated that RiboGen can generate RNA structures that are chemically plausible, as evidenced by the distributions of key geometric parameters including dihedral angles and ribose puckering. Our model outperforms previous approaches in self-consistency evaluation (scTM score) across a wide range of sequence lengths, particularly for longer RNAs. Our early results demonstrate that the generation of RiboGen provides a competitive and efficient RNA design workflow, suggesting sequence-structure co-generation to be a strong approach for RNA modeling. As the field of RNA design continues to grow in importance for therapeutic and biotechnology applications, we believe that generative models like RiboGen will become increasingly valuable tools for exploring and engineering RNA.

\label{gen_inst}

\section*{Acknowledgments}
This research was made possible through the support of the Eleven Eleven Foundation, the Center for Bits and Atoms, and the MIT Media Lab Consortium

\bibliography{iclr2025_conference}
\bibliographystyle{iclr2025_conference}

\newpage
\appendix
\section{Appendix}

\subsection{Data Distribution and Data Balancing}\label{subsec:data_balancing}
\label{apx:data}

\begin{figure}[h]
    \centering
    \includegraphics[width=0.7\linewidth]{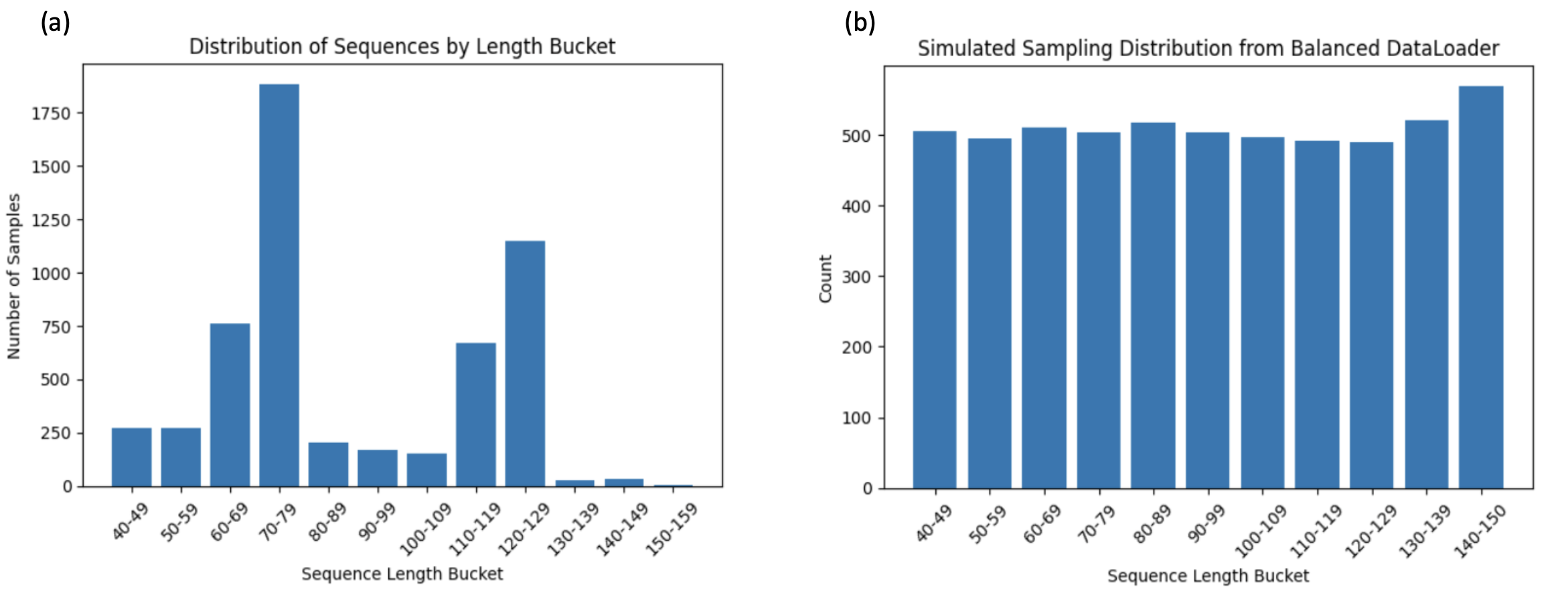}
    \caption{\textbf{Distribution of Sequence Lengths of RNAs in Training Dataset, by the buckets, before and after balancing:} (a) the original distribution of RNA sequences in the training dataset, categorized by length buckets of 10 nucleotides each. The distribution exhibits significant imbalance, with pronounced peaks at 70-79 nucleotides and 120-129 nucleotides, likely corresponding to over-represented tRNA and rRNA classes. In contrast, sequences in the 80-109 range and those longer than 130 nucleotides are substantially underrepresented, with fewer than 200 samples in some buckets. (b) our balanced sampling implementation on the training distribution results in all length buckets are uniformly sampled with approximately 500 sequences per bucket during training. This uniform distribution ensures that the model receives equal exposure to RNA sequences across the entire length spectrum from 40 to 150 nucleotides.}
    \label{fig:rna-bycket-seq-len}
\end{figure}

In RNA sequence analysis, certain RNA types (such as tRNA, rRNA) are over-represented in the different datasets. As shown in Figure \ref{fig:rna-bycket-seq-len}, our original dataset exhibited significant length imbalances, with pronounced peaks at specific length ranges (70-79 and 120-129 nucleotides). This imbalance led to biased model performance, where prediction accuracy was significantly higher for over-represented length ranges but suffered for underrepresented ones.
To address this issue, we implemented a length-balanced sampling in the dataset class that ensures uniform representation of RNA sequences across the entire length spectrum during training without changing the original dataset. Our algorithm divides RNA data into length buckets, with range of 10 per bucket; 40-49, 50-59, etc. (except of the last bucket which contains 140-150) During training a random length bucket is chosen uniformly, and out of it a random datum from this bucket is sampled. It dynamically balances the dataset during training and allows the model to see all available data while preventing over-represented lengths from dominating the training process. This balancing technique led to significant improvements in the model's performance across all length ranges, particularly for previously underrepresented sequences. 
\begin{algorithm}
\caption{Length-Balanced RNA Sequence Sampling}
\label{alg:length_balanced_sampling}
\begin{algorithmic}[1]
\REQUIRE Dataset $D$ containing RNA sequences with lengths $L \in [40, 150]$
\ENSURE Uniformly sampled sequence across all length ranges

\STATE \textbf{Preprocessing:}
\STATE \hspace{1em} Group sequences into buckets $B$ by length range (40--49, 50--59, ..., 140--150)
\FOR{each sequence $s \in D$}
    \STATE Determine length $l$ of $s$
    \STATE Assign $s$ to bucket $B[\lfloor l/10 \rfloor \times 10]$
\ENDFOR

\STATE \textbf{Sampling:}
\STATE Select target length range $t$ uniformly at random from available buckets
    \STATE Return a randomly selected sequence from $B[t]$

\end{algorithmic}
\end{algorithm}

\end{document}

%% file: math_commands.tex
\usepackage{amsmath,amsfonts,bm}

\def\eqref#1{equation~\ref{#1}}

\def\1{\bm{1}}

\DeclareMathAlphabet{\mathsfit}{\encodingdefault}{\sfdefault}{m}{sl}
\SetMathAlphabet{\mathsfit}{bold}{\encodingdefault}{\sfdefault}{bx}{n}

